\documentclass[preprint2]{aastex}

\received{}
\revised{}
\accepted{}
\ccc{}
\cpright{}{}

\slugcomment{Submitted to ApJ}

\shorttitle{MAY 12 1997 CME EVENT: I.}
\shortauthors{TITOV et al.}

\newcommand{\bm}[1]{ \mbox{\boldmath{$#1$}} }
\newcommand{\tA}{\tau}
\newcommand{\nt}{\bm{\nabla}_{\rm t}}
\newcommand{\Ht}{$\bm{\rangle\!\!\!-\!\!\!\langle}$}
 
\begin{document}

\title{MAY 12 1997 CME EVENT: I. A SIMPLIFIED MODEL OF THE PRE-ERUPTIVE MAGNETIC STRUCTURE} 


\author{V. S. Titov, Z. Mikic, J. A. Linker, and R. Lionello}
\affil{Science Applications International Corporation, 10260 Campus
Point Drive, San Diego, California 92121-1578} 
\email{titovv@saic.com
}





\begin{abstract}
A simple model of the coronal magnetic field prior to the
CME eruption on May 12 1997 is developed.
First, the magnetic field is constructed by superimposing a large-scale
background field and a localized bipolar field to model the active
region (AR) in the current-free approximation.
The background field is determined from the normal component of the
observed photospheric magnetic field averaged over the longitude of
the Sun.
The AR field is modeled with the help of a subphotospheric dipole
whose strength, location, and orientation are optimized to fit the
magnetic field obtained from an MDI magnetogram.
Second, this potential configuration is quasi-statically sheared by photospheric vortex motions applied to two flux concentrations of the AR.
Third, the resulting force-free field is then evolved by canceling the photospheric magnetic flux with the help of an appropriate tangential electric field applied to the central part of the AR.

To understand the structure of the modeled configuration, we use the field line mapping technique by generalizing it to spherical geometry.
It is demonstrated that the initial potential configuration contains a hyperbolic flux tube (HFT) which is a union of two intersecting quasi-separatrix layers.
This HFT provides a partition of the closed magnetic flux between the AR and the global solar magnetic field.
Such a partition is approximate since the entire flux distribution is perfectly continuous.
The vortex motions applied to the AR interlock the field lines in the coronal volume to form additionally two new HFTs pinched into thin current layers.
Reconnection in these current layers helps to redistribute the magnetic flux and current within the AR in the flux-cancellation phase.
In this phase, a magnetic flux rope is formed together with a bald patch separatrix surface wrapping around the rope.
Other important implications of the identified structural features of the modeled configuration are also discussed.
\end{abstract}


\section{INTRODUCTION}

The May 12 1997 CME occurred during solar minimum in an isolated active region near disk center.
Since the solar magnetic field during this period had minimum complexity, this event is especially favorable for detailed study.
Therefore, it was selected by several working groups (SHINE, CISM and
MURI) for modeling and analysis.
Several observational features of the event have been reported \citep{
Hudson1998, Plunkett1998, Sterling2001, Thompson1998, Webb2000,
Gopalswamy2002, Lundquist2003, Arge2004, Li2004, Liu2004, Attrill2006, Attrill2007}, which has
provided a good base for creating a magnetohydrodynamic (MHD) model of
the event.
For the latter, we have used the following three-step approach.

First, a potential magnetic field is determined from the observed
photospheric distribution of the normal field component.
Second, the configuration is sheared quasi-statically by photospheric vortex motions applied to the opposite polarities of the active region (AR).
Third, an appropriate tangential electric field is applied then to the central part of the AR to cancel half of the photospheric magnetic flux.
This third step is motivated by the observed flux-canceling   properties of the photospheric motions in the event \citep{Li2004, Ambastha2001}.
Theoretically, the third step is intended first to form a flux rope and then to destabilize it by detaching the rope field from the photospheric one, which has eventually to produce its eruption into the corona and heliosphere.
This complete modeling program has begun by our group and it will be described in forthcoming papers.
Preliminary results for these simulations have been presented by \citet{LinkeretalFAGU2006}.
In this first paper, we consider a simplified model of the magnetic field and its structure at each of the above three steps in order to identify the generic structural features of the configuration prior to its eruption.
By the term  ``generic" we mean those features of the structure that are likely to remain in more accurate models of the AR magnetic field  (AR8038 at N21W08).

Observationally, this field  had an apparently simple bipolar structure \citep{Liu2004} dominating over a more complicated but distant and weak surrounding field.
As a starting point, it is numerically extrapolated in the potential approximation from the synthetic magnetogram obtained by merging an MDI magnetogram with the corresponding synoptic map.  
From the methodological point of view, it is sensible, however, to begin with a simpler model by approximating the numerical potential field by a sum of two components, an azimuthally averaged large-scale magnetic field plus a bipolar field of the active region itself.
This model, comprising essentially two bipolar field components with different scales and orientations, has a deceptive simplicity.
It turns out that the structure of such a field is nontrivial even at the initial current-free state, and it becomes even more complicated at the shearing and flux-canceling phases, features that have important implications for understanding the event.

The analysis of this structure prior to eruption is a major goal of the present paper.
The paper is organized in the following way: the construction of the field model is described in sections \ref{s:ipmf} and \ref{s:mhd_ev}; in section \ref{s:flcsg}, the so-called squashing factor is generalized to the case of spherical geometry in order to investigate in section \ref{s:sif} the structure of the modeled field.
Section \ref{s:imp} describes the relationship between field line connectivity and the current distribution.
Our results are summarized in section \ref{s:s}.
Some details of the initial potential field model are provided in the Appendix.

\section{THE INITIAL POTENTIAL MAGNETIC FIELD} 
	\label{s:ipmf}

Although the computation of potential magnetic fields is routine, the investigation of the field structure and its impact on the subsequent MHD evolution is a novel approach.
For this reason, we prefer to idealize  our initial field ${\bm B}_{0}$ as much as possible by keeping only the most essential properties of the potential field computed from magnetogram data.
Since our primary interest lies in exploring the eruption process, the
initial field ${\bm B}_{0}$ ought to incorporate at the very least the
large-scale coronal field ${\bm B}_{\sun}$, which we call the
background field, and the field ${\bm B}_{\rm AR}$ of the active
region itself.
Thus, the resulting field can be written as
\begin{eqnarray}
	{\bm B}_{0} = {\bm B}_{\sun} + {\bm B}_{\rm AR}
	= - {\bm \nabla} F_{\sun} - {\bm \nabla} F_{\rm AR},
		\label{B0}
\end{eqnarray}
where $F_{\sun}$ and $F_{\rm AR}$ are harmonic potentials of the
magnetic field.
It should be noted that although both fields have a similar bipolar
character, they have in general different orientations and length
scales: the background field varies on the scale of the solar radius
$R_{\sun}$, which greatly exceeds the size of the active region $L$.

\subsection{Background field ${\bm B}_{\sun}$}

\begin{figure}[htbp]
\epsscale{1.0}
\plotone{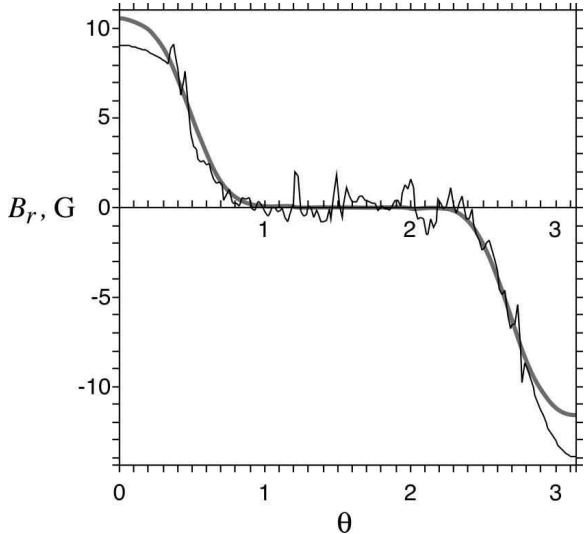}\\
\caption{The latitudinal distribution of the photospheric radial
magnetic field after averaging over longitude, together with its smoothed
analytical approximation (thin solid and thick grey-shaded lines,
respectively).
	\label{f1} }
\end{figure}

For the sake of simplicity, we also assume that ${\bm B}_{\sun}$ and,
in particular, its photospheric radial component $\left. B_{r\sun}
\right|_{r=R_{\sun}}$, is axisymmetric.
The latter can be approximated by averaging the observed
$B_{r}(R_{\sun},\theta, \phi)$ over the longitude $\phi$ and then by
smoothing and fitting it to the sum
\begin{eqnarray}
	\left. B_{r\sun} \right|_{r=R_{\sun}} = \sum_{n=n_{0}}^{N}
	c_{n}\, \cos^{n}\theta ,  
		\label{Brbp}
\end{eqnarray}
where the coefficients $c_{n}$ are determined by a least square
minimization.
Imposing additionally the flux-conservation condition and sacrificing
the accuracy of fitting at the poles, where the observational data
are not reliable because of the projection effect, we have found that
equation (\ref{Brbp}) provides at $n_{0}=5$ and $N=11$ quite a good
approximation to the large-scale field of the Sun (Figure~\ref{f1}).
%

Under these assumptions, the potential of the background field can
be expressed in terms of spherical functions as
\begin{eqnarray}
	F_{\sun} = \sum_{n=1}^{N} (C^{-}_{n} r^{-n-1} +C^{+}_{n} r^{n})
	P_{n}(\cos\theta) ,
		\label{Fbg}
\end{eqnarray} 
where $P_{n}(\cos\theta)$ is a Legendre polynomial  of $n$-th order. 
The terms with $n=0$ are omitted here because the total magnetic flux
must be zero and $F_{\sun}$ is defined up to an arbitrary constant.

The unknown coefficients $C^{\mp}_{n}$ in this expression are
determined by linear algebraic equations derived from the boundary or
asymptotic conditions.
The first number $N$ of such equations can be obtained by equating the
corresponding linear combinations of $C^{\mp}_{k}$, $k=1,\ldots,n$, at
$\cos^{n}\theta$ in
\begin{eqnarray}
	\left. B_{r\sun} \right|_{r=R_{\sun}}  = 
	        -\left. \frac{\partial F_{\sun}}
			           {\partial r} \right|_{r=R_{\sun}}
	\label{BC1}
\end{eqnarray}
to either 0 or $c_{n}$ depending on $n$ (see equation (\ref{Brbp})): the first and second case corresponds to $n=1,\ldots,n_{0}-1$ and $n=n_{0},\ldots, N$, respectively.
These equations are rather cumbersome, but they can easily be retrieved, so we do not present them here explicitly.

The form of the remaining $N$ equations for $C^{\mp}_{n}$ depends on what kind of behavior at large radii $r$ we would like to have for ${\bm
B}_{\sun}$.
The simplest choice is to require $\left. B_{\sun} \right|_{r
\rightarrow \infty} \rightarrow 0$,  for which the
corresponding coefficient equations are simply
\begin{eqnarray}
	C^{+}_{n}=0, \quad n=1,\ldots,N ,
		\label{BC2_1} 
\end{eqnarray} 
-- and this is one of the two cases that we have considered.

However, for modeling the solar wind effect, the second choice of boundary condition, namely, 
\begin{eqnarray*}
	\left. F_{\sun} \right|_{r=R_{\rm ss}} = 0 
\end{eqnarray*}
with the source-surface radius $R_{\rm ss}$ equal, say, to $ 2.5\, R_{\sun}$, is also of interest.
According to equation (\ref{Fbg}), this yields
\begin{eqnarray}
	R_{\rm ss}^{-n-1}\, C^{-}_{n} + R_{\rm ss}^{n} C^{+}_{n} = 0,
			\quad n=1,\ldots,N,
		\label{BC2_2}
\end{eqnarray}
which reduces to equation (\ref{BC2_1}) at $R_{\rm ss} \rightarrow \infty$, as expected.
The magnetic field under this condition becomes radial at the sphere of radius $R_{\rm ss}$, and the corresponding approach is called the potential field source surface (PFSS) model \citep{Altschuler1969, Schatten1969}.
This type of solution models an open field structure sustained by the solar wind, which makes it possible to estimate its impact on the coupling of ${\bm B}_{\sun}$ and ${\bm B}_{\rm AR}$.

For the analysis of the field line structure in both configurations,
it is useful also to have an explicit expression of the magnetic
flux function $\Psi_{\sun}(r,\theta)$, which in our case is 
\begin{eqnarray}
	&& \Psi_{\sun} = \sum_{n=1}^{N} \left[ \left( 1+\frac{1}{n} \right)
		C^{-}_{n} r^{-n} -C^{+}_{n} r^{n+1} \right] \times
			\nonumber \\
	&& \qquad\qquad \left[\cos\theta P_{n}(\cos\theta) - P_{n+1}(\cos\theta) \right] .
		\label{Psi}
\end{eqnarray}
The contours of $\Psi_{\sun}$ in the planes $\phi={\rm const.}$ define
the field lines of ${\bm B}_{\sun}$, since
\begin{eqnarray}
	{\bm B}_{\sun} = {\bm \nabla}\Psi_{\sun} \times 
	{\bm \nabla}\phi .
	\label{PsPh}
\end{eqnarray}

We chose $R_{\sun}$ to be a unit length scale in the problem, so that
the coefficients of both type of solutions are calculated from the above
equations at $R_{\sun}=1$. Table \ref{t1} in appendix \ref{s:pars} presents their numerical values and shows that the coefficients $C^{-}_{n}$ of these two
solutions are not very different, and the coefficients $C^{+}_{n}$ of
the second solution are rather small.
This suggests that the solar wind does not much affect the magnetic
field in the lower part of the corona, in agreement with our
study of the entire magnetic field~${\bm B}_{0}$ described below.
%
%
\begin{figure}[htbp]
\epsscale{0.8}
\plotone{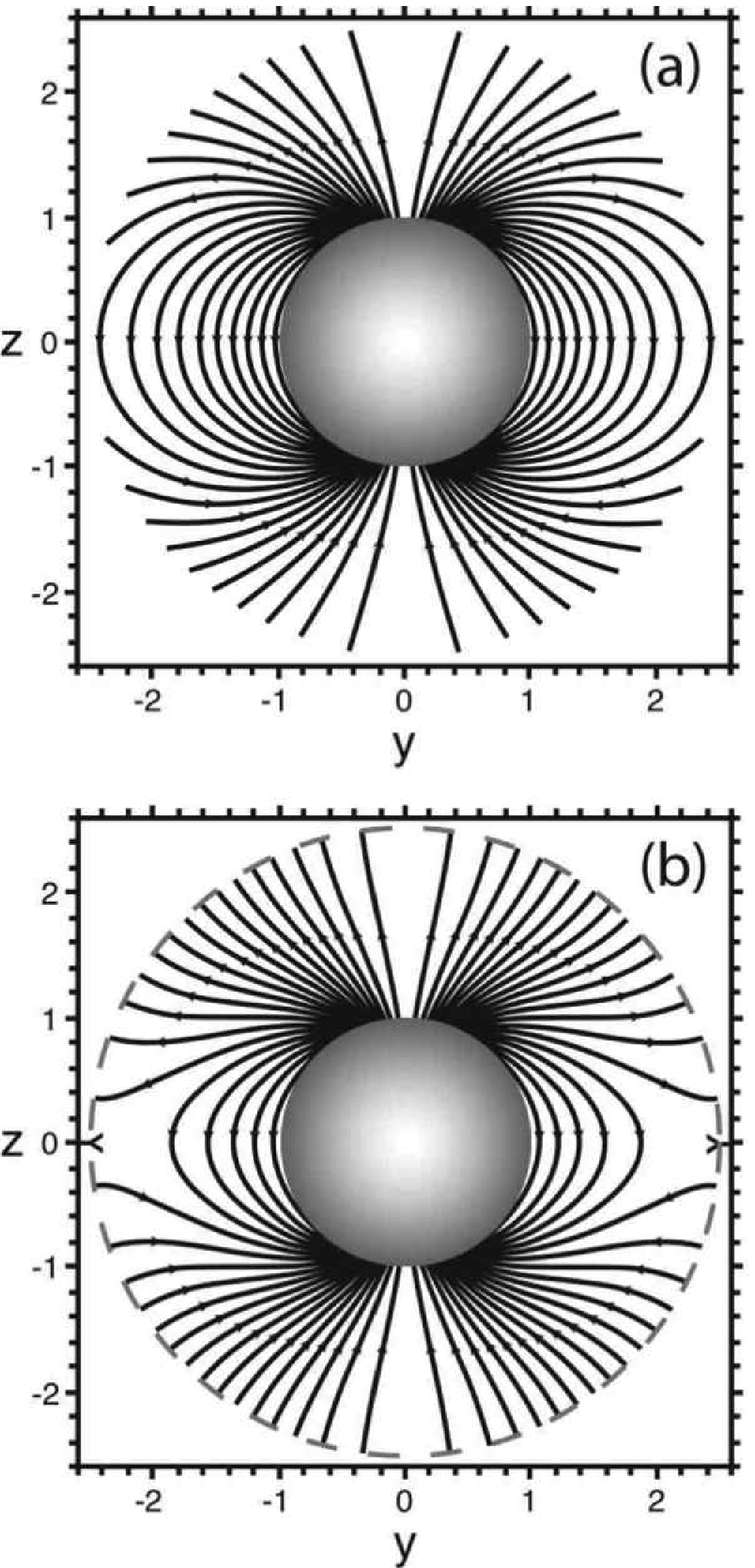}
\caption{Magnetic field lines of the axisymmetric background potential
field ${\bm B}_{\sun}$ in the plane $x=0$ at $r \le R_{\rm ss} = 2.5
R_{\sun}$ for two different boundary conditions (a) $\left. B_{\sun}
\right|_{r\rightarrow\infty}
\rightarrow 0$ and  (b) $\left. F_{\sun}\right|_{r=R_{\rm ss}}=0$.
 The null points of the second configuration are indicated by Y
 letters turned by $\pm90\degr$.
	\label{f2} }
\end{figure}
It is also in agreement with the structure of the field lines plotted
as equidistant contours of the flux function $\Psi_{\sun}$ in the
plane $x=0$ of a Cartesian system of coordinates $(x,y,z)$.
Figure \ref{f2} clearly demonstrates that both structures are nearly
the same everywhere accept for $r \ga 1.3 R_{\sun}$ at equatorial
latitudes.
As expected, the source-surface configuration has a null line along the equator of the sphere $r=R_{\rm ss}$ to accommodate the direction reversal of the open field. 

\subsection{Active-region field ${\bm B}_{\rm AR}$}

The normal component of the photospheric magnetic field in our
AR has a well-pronounced bipolar character.
It is natural therefore to approximate the corresponding potential
field with the help of a fictitious subphotospheric dipole, whose
potential is
\begin{eqnarray}
	F_{\rm AR} = \frac{{\bm m}_{\rm d} \cdot ({\bm r} - {\bm r}_{\rm d})} 
		                   {|{\bm r} - {\bm r}_{\rm d}|^{3}}  .
	\label{Far}
\end{eqnarray}
This analytical expression was used to fit the numerical potential
field solution ${\bm B}_{\rm num}$ computed from the original
magnetogram data.
We have optimized the location of the dipole ${\bm r}_{\rm d}$ and its
moment ${\bm m}_{\rm d}$ to minimize the difference between $B_{r\sun}
+ B_{r\rm AR}$ and $B_{r\,{\rm num}}$.
%
%
\begin{figure}[htbp]
\epsscale{1.0}
\plotone{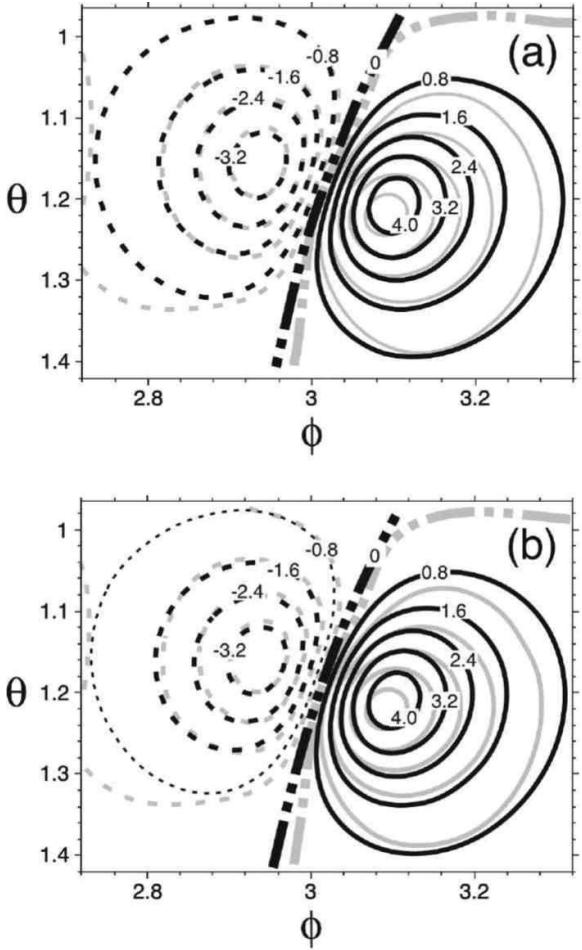}
\caption{Contours of $ B_{r\,{\rm AR}} $ and
$B_{r\,{\rm num}} - B_{r\sun}$ (grey-shaded) at $r=r_{\rm opt}$ in
the $(\phi,\theta)$-plane for two different
boundary conditions (a) $\left. B_{\sun} \right|_{r\rightarrow\infty}
\rightarrow 0$ and  (b) $\left. F_{\sun}\right|_{r=R_{\rm ss}}=0$.
	\label{f3} }
\end{figure}
%
More precisely, ${\bm r}_{\rm d}$ and ${\bm m}_{\rm d}$ are fitted to
minimize the sum of $(B_{r\sun} + B_{r\,\rm AR}-B_{r\,{\rm num}})^2$
evaluated at 14 points which were randomly scattered over the AR on the sphere $r =r_{\rm opt}$.
The radius $r_{\rm opt}$ is arbitrary to a certain extent, and its
value was chosen equal to $1.12 R_{\sun}$ to filter out
magnetic inhomogeneities of the size $\la 0.1 R_{\sun}$ in the lower
part of the solar atmosphere.

In spite of its simplicity, the described analytical model provides a
relatively good approximation of the numerical potential field at $r
=r_{\rm opt}$ (see Figure \ref{f3}).
The model becomes even more accurate at larger $r$, because higher-order harmonics describing the small-scale inhomogeneities in the
numerical solution decay faster with $r$ than the lower-order ones.
 
Table \ref{t2} in appendix \ref{s:pars} presents the resulting
optimized parameters of the dipole in spherical coordinates for both
types of the above boundary conditions.
It shows that the magnetic moments $m_{r}$ of these two types of
solutions differ less than $2.4\%$ from each other, while the
differences between other parameters are even smaller.
This is consistent with the above conclusion that the solar wind does
not greatly influence the magnetic structure in the low corona.

\section{QUASI-STATIC MHD EVOLUTION OF THE MAGNETIC CONFIGURATION}
	\label{s:mhd_ev}
	
The above potential field with the boundary condition $\left. B_{\sun} \right|_{r \rightarrow \infty} \rightarrow 0$ is used as the initial configuration for our simplified MHD model of the AR evolution prior to eruption.
This field is not intended to describe the coronal field at any instant of its evolution, but rather to serve as a convenient starting point for further modeling.
To energize this configuration, it is quasi-statically sheared by the photospheric flows that preserve the initial photospheric radial component of magnetic field.
This step produces a stressed configuration with a nearly force-free magnetic field.
Its subsequent evolution is driven by imposing a special photospheric distribution of the tangential electric field that models the observed flux-cancellation process.
	
\subsection{Shearing/twisting of the magnetic flux spots}
	\label{s:sh}

The initial potential field ${\bm B}_{0}$ is sheared by photospheric flows such that its footpoints are rotated in the flux spots of the AR in opposite directions by following approximately the contours of the initial photospheric distribution of the radial field component $B_{0r_{\sun}} (\phi,\theta) \equiv \left. B_{0r} \right|_{r=R_{\sun}}$.
The corresponding shearing flow is described by
\begin{eqnarray}
  {\bm v}_{\rm sh}(\phi,\theta) = \frac{1}{B_{0r_{\sun}}}
     \nt \times \left( f B_{0r_{\sun}}^{3}
     \hat{\bm r}\right) ,
	\label{v_sh}
\end{eqnarray}
where ${\bm \nabla}_{\rm t}$ stands for the gradient operator tangential to the photosphere.
This velocity field vanishes at the PIL defined by $B_{0r_{\sun}} (\phi,\theta)=0$ and preserves the initial distribution of $B_{0r_{\sun}} $, since ${\bm \nabla}_{\rm t} {\bm \cdot} \left(B_{0r_{\sun}} {\bm v}_{\rm sh} \right) = 0$.

The function
\begin{eqnarray}
   f = a(t)\, \exp \left[ - \delta^{2}  \left( \frac{\sin^{2}\gamma} {w_{\theta}^{2}} + \frac{\cos^{2}\gamma} {w_{\phi}^{2}}\right)\right] 
	\label{f}
\end{eqnarray}
provides an oval mask concentrating the shear in the AR near the PIL, where $a(t)$ is an amplitude linearly ramping until $a(t)=0.025\, R_{\sun}/\tA$ is reached at $t = 0.5\, \tA$.
After that $a(t)$ remains constant until $t= 3.0\, \tA$, when it drops linearly to zero at $t= 3.05\, \tA$ and remains zero from there on.
The time unit $\tA$ must be large enough to guarantee quasi-static evolution of the whole magnetic configuration.
Specifically, it was chosen to be $\sim 1800$ times the Alfv\'{e}n time in the AR, and 4 times the Alfv\'{e}n time in the polar field region.

\begin{figure}[htbp]
\epsscale{1.0}
\plotone{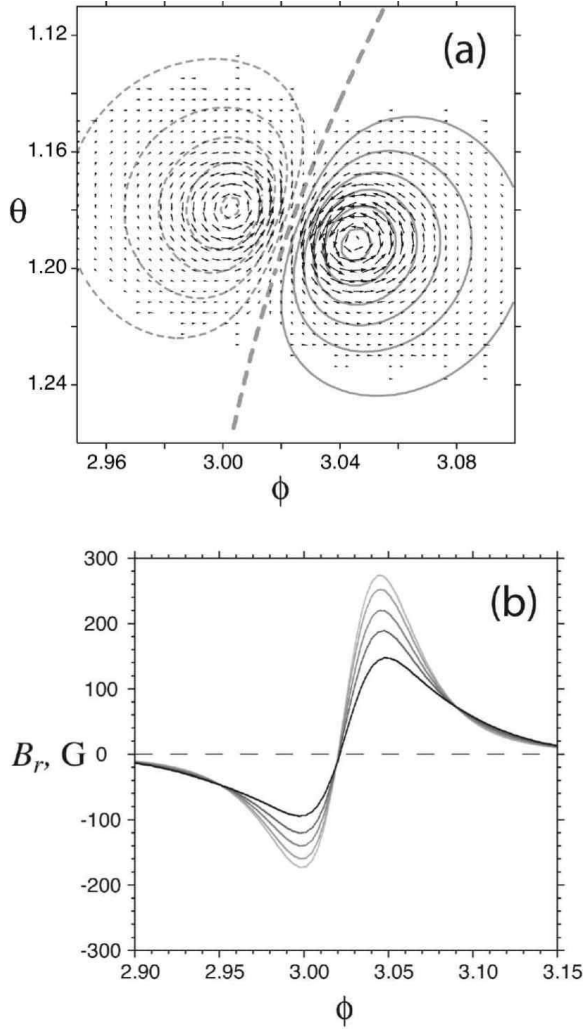}
\caption{ (a) The vector field of the photospheric velocity in the shear/twist phase and the respective contours of $\left. B_{r} \right|_{r=R_{\sun}} = 44\, i \;{\rm G},\: i=-5,\ldots,6$; solid and dashed lines correspond to positive and negative values, respectively, while the thick dashed line represents the PIL. (b) The longitudinal profiles of $\left. B_{r} \right|_{r=R_{\sun}}$ in the AR at the middle latitude $\theta_{*}=1.187$ for five equidistant time moments: the darkness of the grey-colored curves is monotonously increased from the initial to the final profile.
	\label{f4} }
\end{figure}

The angle variables are measured hereafter in radians and defined as 
\begin{eqnarray}
  \gamma & = & \gamma_{0} + \arctan\left(\theta_{0} - \theta, \phi - \phi_{0} \right) ,
	\label{gamma} \\
	\delta & = & \arccos \frac{{\bm R} {\bm \cdot} {\bm R}_{0}}{R_{\sun}^{2}} ,
	\label{delta}
\end{eqnarray}
where $\gamma_{0} \approx -1.396$ is a tilt of the oval, ${\bm R}$ and ${\bm R}_{0}$ are the radius-vectors of the photospheric points with longitudes $\phi$ and $\phi_{0} = 3.026$ and latitudes $\theta$ and $\theta_{0} = 1.184$, respectively.
The parameters $w_{\theta}$ and $w_{\phi}$ controlling the sizes of the oval mask are chosen to be $w_{\theta} = 0.06$ and $w_{\phi} = 0.02$.
The resulting photospheric velocity field determined by equations (\ref{v_sh})--(\ref{delta}) is shown for the described set of parameters in Figure \ref{f4}a.

\subsection{Model of the flux cancellation}
	\label{s:fc}

The process of flux cancellation is modeled by imposing at the photosphere the following tangential electric field
\begin{eqnarray}
  {\bm E}_{\rm t} = \nt \bm{\times} \left( g \psi \hat{\bm r} \right) ,
	\label{Et}
\end{eqnarray}
where the flux function $\psi$ is determined from the two-dimensional Poisson equation on the sphere $r=R_{\sun}$ such that
\begin{eqnarray}
   \nt^{2} \psi = B_{0r_{\sun}} .
\end{eqnarray}
The choice of a spatially uniform factor $g$ provides a simple uniform reduction of $B_{0r_{\sun}}$.
However, the flux function $\psi$ and the related electric field ${\bm E}_{\rm t}$ decay in this case too slowly with distance from the AR compared to the respective magnetic field.
This leads unfortunately to the undesirable formation of a strong boundary layer covering a substantial area outside of the AR.
To localize ${\bm E}_{\rm t}$ within the AR, the mask function 
\begin{eqnarray}
   g = c(t)\, \exp\left[ -
     \frac{\left(\phi-\phi_{*}\right)^{2} + \left(\theta-\theta_{*}\right)^{2}}{w^{2}} \right] 
	\label{g}
\end{eqnarray}
is used with $\phi_{*} = 3.022$ and $\theta_{*} = 1.187$ representing an approximate position of the AR center, and $w=0.065$ determining the extent of the electric-field localization.
The amplitude $c(t)$ here is linearly increased from $t = 3.1\, \tA$ until some $c(t)=c_{\rm max}$ is reached at $t  = 3.3\, \tA$.
After that, $c(t)$ remains constant until $t= 7.5\, \tA$, when it drops linearly to zero at $t= 7.7 \, \tA$ and remains zero from there on.
The constant $c_{\rm max}$ is chosen from the requirement to reduce the AR flux approximately to a half of its initial value in the specified interval of time.
The resulting reduction of the photospheric normal magnetic field in the described flux-cancellation model is depicted in Figure \ref{f4}b.

\section{FIELD LINE CONNECTIVITY IN SPHERICAL GEOMETRY}
	\label{s:flcsg}

As we will see, our initial configuration has a simple topology such that no magnetic null points (NPs) or bald patches (BPs) are present in the modeling AR.
BPs are segments of the PIL at which field lines are concave towards the corona \citep{Titov1993}.
An exception is the above-mentioned null line at the source surface
$r=R_{\rm ss}$, which accommodates the direction reversal of the
open field at the equator in the PFSS model.
This null line, however, is outside of the AR, so its
structure may have only geometrical features relevant for
understanding the eruption process under study.
The features of major interest are quasi-separatrix layers (QSLs,
\citet{Priest1995}) and, particularly their combination in
the form of two self-intersecting QSLs, the so-called hyperbolic flux
tubes (HFTs, \citet{Titov2002}).
This is because HFTs have properties favorable for developing current sheets and reconnection \citep{Titov2003, Galsgaard2003, Aulanier2005}, even if the respective configuration is free of NPs and BPs.
The properties of such structural features and their manifestation in observed events are recently reviewed in detail by \citet{Demoulin2005, Demoulin2006}.

These features are determined with the help of a point-wise analysis of the magnetic field line mapping.
For small ARs, where the photosphere can be approximated by a plane, it is convenient to define this mapping in a Cartesian system of coordinates $(x,y)$.
Such coordinates allow us not only to determine the location of the footpoints but also to measure the distances between them in the $x$ and $y$ directions.
The mapping $(X(x,y), Y(x,y))$ defines the connection between one
footpoint $(x,y)$ and another $(X,Y)$ by a magnetic field line.
Notice that this mapping is defined only for closed field lines, while its generalization for open field lines requires an introduction of an upper boundary surface (see below).
The local properties of the field line mapping are described by the
Jacobian matrix
 \begin{eqnarray}
 D
  = \left( \begin{array}{cc}
 {\partial X \over \partial x}
    & {\partial X \over \partial y}\\
 {\partial Y \over \partial x}
    & {\partial Y \over \partial y}
           \end{array} 
    \right)
 \equiv \left( \begin{array}{cc}
    a    & b\\
    c    & d
           \end{array}
    \right) ,
  \label{D}
 \end{eqnarray}
whose elements, however, are not invariant with respect to the direction of mapping.
Therefore, they cannot be used directly for characterizing the field
line connectivity and for detecting QSLs in configurations.
One can prove that a natural quantity which satisfies this requirement in the case of a plane photospheric boundary is expressed in terms of the elements of $D$ as \citep{Titov2002}
\begin{eqnarray}
Q  = N^{2}/|\Delta|, 
        \label{Q}
\end{eqnarray}
where
\begin{eqnarray}
	N^2       =  a^2 + b^2 + c^2 + d^2  
		\label{N}
\end{eqnarray}
is the so-called norm squared \citep{Priest1995} and
\begin{eqnarray}
	\Delta  &=&  a d - b c
		\label{Dlt}
\end{eqnarray}
is the Jacobian of the mapping.
To explain the origin of $Q$, note that the Jacobian matrix represents
a linearized field line mapping in the vicinity of a given footpoint.
Therefore, any infinitesimal circle centered at such a point will be
mapped along the neighboring field lines into an infinitesimal ellipse
at the other footpoint.
It can be shown that the value $Q/2+\sqrt{Q^2/4-1}$ determines
the aspect ratio of such an ellipse [see the details in \citep{Titov2007a}].
It characterizes the degree of squashing of the corresponding infinitesimal magnetic flux tubes and turns simply into $Q$ for $Q \gg 2$.
Since QSLs are determined by large values of $Q$, we will neglect
the slight difference between $Q$ and the above exact value of squashing degree by calling $Q$ itself the squashing degree or squashing factor \citep{Titov2002}.

Another independent geometrical quantity invariant to the direction of
mapping (up to the sign) is even simpler than $Q$.
It determines the degree of expansion or contraction of  infinitesimal flux tubes and, due to the flux conservation in them, can be expressed in terms of the ratio of the normal field components at the connected footpoints as \citep{Titov2002}
\begin{eqnarray}
	K \equiv \log |\Delta| = \log |B_{n}/B^{*}_{n}| .
	\label{K}
\end{eqnarray}
Here $B^{*}_{n}$ and $B_{n}$ are normal components of the magnetic field at the conjugate ``target'' and ``launch'' footpoints, $(X(x,y),Y(x,y))$ and $(x,y)$, respectively.
In practice, the numerical calculation of $\Delta$ through this ratio
is more precise than that given by equation (\ref{Dlt}) and therefore it should be used for computing $Q$ in equation (\ref{Q}) as well.

The quantities $Q$ and $K$ provide a valuable tool for investigating
the geometrical structure of coronal magnetic fields.
For example, they made it possible to convincingly demonstrate that
HFTs are a generalization of separator field lines to the case of
topologically simple magnetic configurations having no NPs or BPs
\citep{Titov2002}.
This is particularly interesting for our study, since HFTs
appear naturally in quadrupole configurations \citep{Titov2002}, which can be thought of as being the superposition of two nested dipole fields---exactly the type of potential field considered in the previous section.

Such a field-line mapping technique can be generalized in covariant form to the case where the magnetic configuration is bounded by curved surfaces of arbitrary shape \citep{Titov2007a}.
Here we will restrict ourselves to spherical boundaries and use a much simpler heuristic approach to derive $Q$ in a global spherical system of coordinates.
Notice first that our coronal magnetic field fills the volume between two spheres of radii $R_{\sun}$ and $R_{\rm ss}$.
For measuring distances between infinitesimally close footpoints, one needs to introduce two-dimensional Euclidean bases, which are tangential to these boundaries at each of their points.
In the traditional spherical system of coordinates with the latitude
$\theta=0$ or $\pi$ at the poles of the Sun, these must be pairs of
orthonormal vectors $({\bm e}_{\phi}, {\bm e}_{\theta})$.
Then the infinitesimal increments of length along ${\bm e}_{\phi}$ and ${\bm e}_{\theta}$ in one of the two polarities are $\delta x = R_{\sun} \sin\theta\, {\rm d} \phi$ and $\delta y = R_{\sun}\, {\rm d} \theta$, respectively.
Let $(\Phi(\phi,\theta), \Theta(\phi,\theta))$ be the vector function representing the field line mapping, then similar increments at the conjugate footpoint are $\delta X = R_{*} \sin\Theta\,  {\rm d} \Phi$ and $\delta Y = R_{*}\,  {\rm d} \Theta$ with either $R_{*}=R_{\sun}$ or $R_{*}=R_{\rm ss}$ depending on whether the conjugate footpoint is located at the photosphere or at the source surface, respectively.
Thus, for the elements of matrix (\ref{D}), we obtain in our case
\begin{eqnarray}
	a = \frac{R_{*}\sin\Theta}{R_{\sun}\sin\theta} 
		\frac{\partial \Phi }{\partial \phi }, \quad
		&& b = \frac{R_{*}\sin\Theta}{R_{\sun}} \frac{\partial \Phi }{\partial
			\theta}, 
		\label{ab} \\
	c = \frac{R_{*}}{R_{\sun}\sin\theta} \frac{\partial \Theta }{\partial
		\phi }, \quad 
		&& d = \frac{R_{*}}{R_{\sun}}\frac{\partial \Theta }{\partial \theta }.
	\label{cd}
\end{eqnarray}
It is these expressions that must be used for computing norm (\ref{N}) in squashing factor (\ref{Q}) generalized for the case of both open and closed magnetic fields in the coronal volume restricted by spherical boundaries.
The value of $\Delta$ as a part of the whole expression for $Q$ is more accurately computed through the ratio of the normal (radial) field components, as already noted before.

The above expressions for $a$ and $c$ have apparent singularities at
the poles, where they actually reduce in the generic case to resolved
indeterminacies with $\Phi(\phi,\theta)$ and $\Theta(\phi,\theta)$
proportional to $\sin \theta$.
Such indeterminacies are due to pole singularities of the global spherical system of coordinates rather than special properties of the field line mapping.
Computationally, however, this requires care in evaluating $Q$ near the poles, for example, by excluding them from numerical grid.
A more accurate approach is to use a covariant definition of $Q$ with two overlapping regular coordinate charts covering the spherical boundaries  \citep{Titov2007a}.

\section{FIELD STRUCTURE}
	\label{s:sif}

The photospheric distributions of $Q$ and $K$ are a two-dimensional imprint of the three-dimensional (3D) coronal field structure and so are very helpful for its analysis.
Since their expressions (\ref{Q})--(\ref{cd}) are applicable to arbitrary field configurations, we have computed representative examples of the $Q$ and $K$ distributions for each of the three stages of magnetic field described above.
In general, if the magnetic configuration has a separatrix surface (SS) due to the presence of an NP in the corona or a BP at the PIL, the field line mapping experiences a jump at the SS footprints and the corresponding Jacobian matrix is not defined there.
Numerical derivatives, however, are estimated as a ratio of finite coordinates differences, so their absolute values are limited.
Thus, numerically, the $Q$ factor is always defined at the SS footprints and its narrow spiky distribution there helps identify NP SSs or BP SSs, if present.
\onecolumn
\begin{figure}[htbp]
\epsscale{0.8}
\plotone{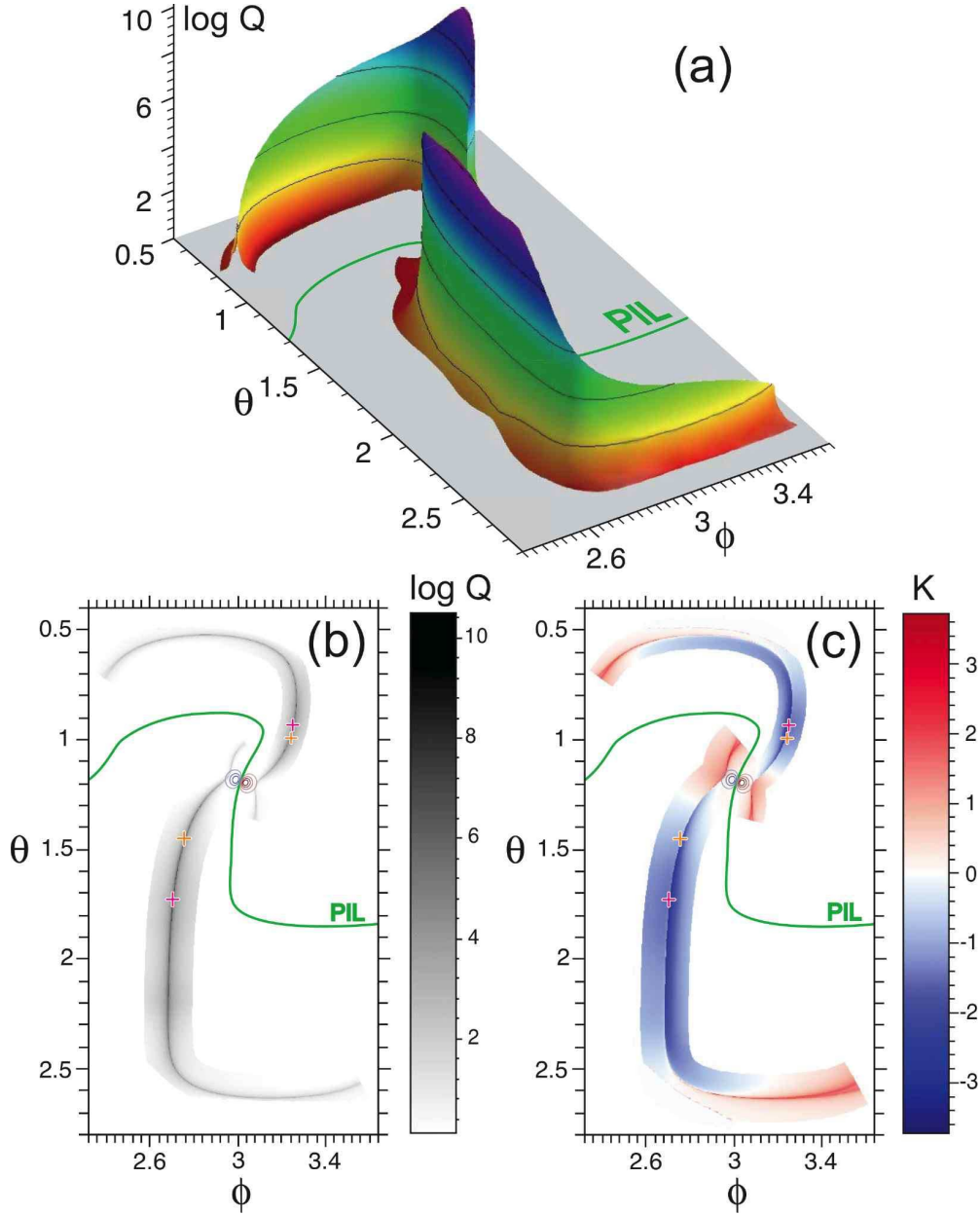}
\caption{Distributions of squashing factor $Q$
(a,b) and expansion-contraction factor $K$ (c) at the HFT footprints.
On top of the grey shaded $Q$ distribution (b) and red-blue shaded
$K$ distribution (c), the polarity inversion line (PIL, green
line), the contours of $\left. B_{r} \right|_{r=R_{\sun}} = 56\, i
\;{\rm G},\:
i=-3,\ldots,4$ (blue and red lines for negative and positive values,
respectively), two points of local minimum of $ \left. B
\right|_{r=R_{\sun}}$ (orange crosses) and footpoints of
quasi-separator (magenta crosses) are superimposed.
Only the parts of the distributions corresponding to the closed field lines are shown.
	\label{f5} }
\end{figure}
\twocolumn
{ \parindent=0pt
On the other hand, large values of $Q$, smoothly distributed on the photosphere in narrow strips, correspond to the footprints of QSLs \citep{Titov2002}.
These general properties of the $Q$ distribution are fully confirmed in the structural analysis of our field model. }

\subsection{The initial potential field}
	\label{s:sipf}

As expected from Section \ref{s:ipmf}, the structure of the initial potential field in the AR turns out to be rather similar for both our models.
Therefore we will describe further only the structure whose background field is determined in the framework of the PFSS model.
The computed $Q$ distribution reveals two very sharp ridges with maxima $\sim 10^{10}$, one for each magnetic polarity (see Figure \ref{f5}a,b).
Each ridge spreads from the respective bipole spot toward the pole region, curving in a sickle-like shape.
They trace the footprints of the QSLs which combine into one flux tube, called an HFT.
The cross-section of the HFT varies in a rather nontrivial way: neglecting some curving and twisting of the HFT, its cross-section can be represented as follows  
\begin{equation}
	\begin{picture}(125,8)(0,0)	
		\put(0,0){
			\includegraphics*[scale=0.18]{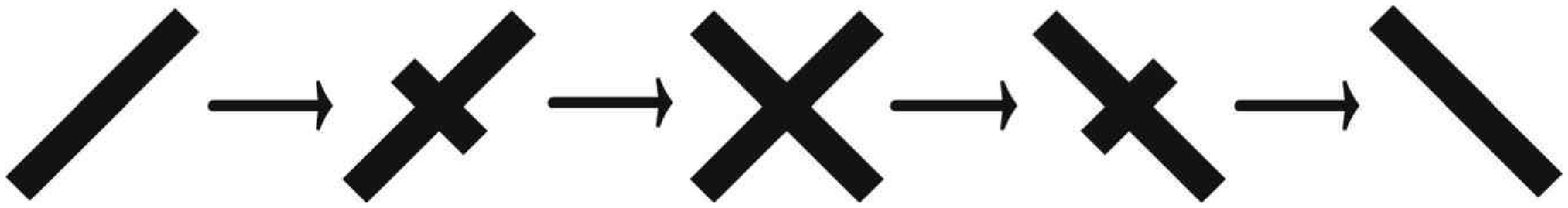}
		}
	\end{picture}.
\end{equation}
This is possible because the cross-section shrinks in the longitudinal direction and stretches in the transverse direction when following the field lines.
The dependence of such a variation on the arc length of the field lines has an exponential character.

\begin{figure}[htbp]
\epsscale{1.0}
\plotone{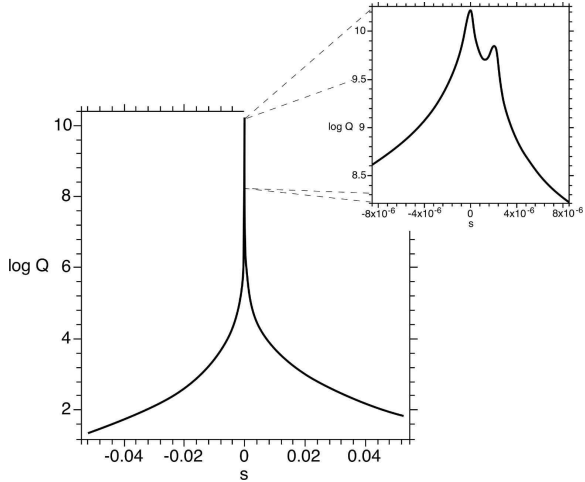}
\caption{Profile of the $\log Q$ distribution in the perpendicular
direction to the HFT footprint; $s$ is the distance in this direction
from one of the footpoints of the quasi-separator $(\phi,\theta)
\approx (3.251, 0.935)$ (the north-east magenta cross in Figures
\ref{f5} (b,c)).
	\label{f6} }
\end{figure}

Because of this, resolving the internal structure of our HFT is numerically challenging.
For this purpose, a curvilinear and nonuniform adaptive mesh with a minimal cell-size of $< 10^{-5}R_{\sun}$ has been developed.
To control the effect of truncation errors, the computation of $Q$ has been performed both with 10 and 12 significant digits by using the computer algebraic system Maple.
The resolved profile of $Q$ across the ridge of the $Q$ distribution
in its highest part is shown in Figure \ref{f6}.
Similar to the case of a quadrupole magnetic field \citep{Aulanier2005}, the $Q$ distribution has two peaks, which are probably related to the presence of two very localized minima of $|{\bm B}|$ at the HFT footprints (see below).
Therefore, strictly speaking, the $Q$ distribution has two sharp ridges closely bordering one another in each of the strips.
However, because this fine-scale feature of the HFT structure is not significant for the results we report here, we neglect it in the subsequent discussion.

There is also a strong similarity between $K$ distributions of the quadrupole \citep{Titov2002} and our configuration.
In both cases, a very steep gradient of $K$ exists  across the ridges of the $Q$ distribution.
On both sides of the ridges in the middle part of the strips, there are areas of negative $K$ (blue colored in Figure \ref{f5}c).
If crossing the strips in these areas outward from the PIL, the $K$ distribution with moderate negative values steepen first into a sharp front of highly negative values.
It is followed then by a sharper front of positive values (red colored but not resolved in Figure \ref{f5}c), which finally drops back to moderate negative values.
If moving along the strips from their middle to their ends, the areas and the fronts of negative $K$ gradually disappear, while the fronts of positive $K$ increase in amplitude and broaden up to the entire width of the strips.

By definition, positive and negative values of $K$ correspond to expansion and contraction, respectively, of elemental flux tubes.
This behavior of the flux tubes is opposite at the conjugate footpoints: if a given flux tube expands toward one footpoint, it contracts toward the other.
Thus,
\onecolumn
\begin{figure}[htbp]
\epsscale{0.8}
\plotone{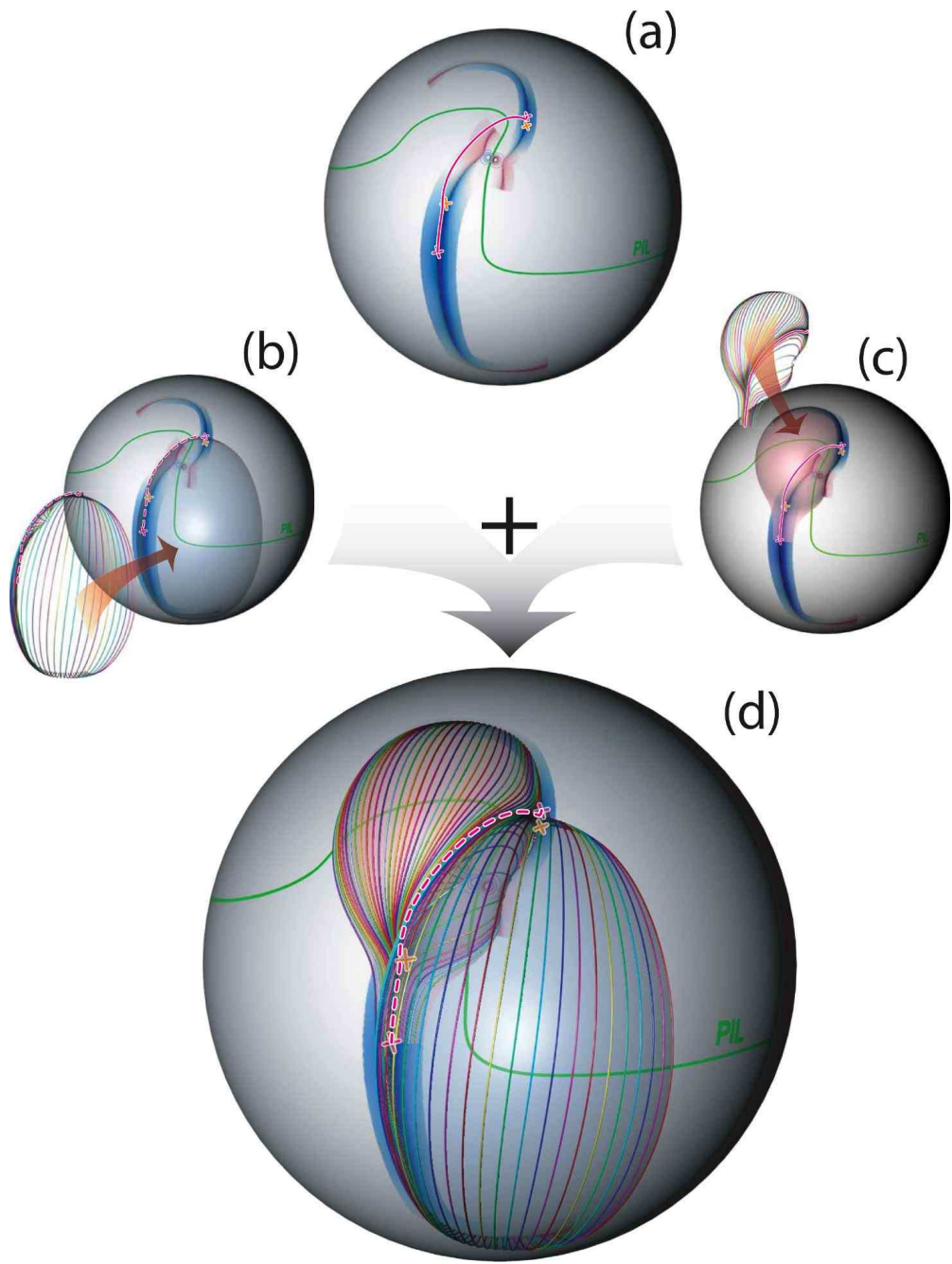}
\caption{Map of the superimposed grey and red-blue shaded,
respectively, $Q$ and $K$ distributions (see Figures \ref{f5}b and
\ref{f5}c) on the solar surface (a); the process of assembling of the quasi-separatrix surfaces (QSSs) from the field lines launched at two ridges (b,c) of the $Q$ distribution to give the entire structure (d).  The intersection of the QSSs determines a quasi-separator field line
(magenta solid (c) or dashed (b,d) line) with the footpoints marked by
magenta crosses.  Orange crosses are local minima of the
photospheric $|{\bm B}|$, and the green line is the PIL.
	\label{f7} }
\end{figure}

\twocolumn
{\parindent=0pt
the conjugate footpoints of the field lines are colored in Figure \ref{f5}c in red and blue of equal intensity.
In the $Q$ distribution (Figure \ref{f5}b), the conjugate footpoints are colored in grey of equal intensity.
In combination with the PIL and contours of the photospheric $B_r$ superimposed on such distributions, this provides a comprehensive graphical representation of the field line connectivity.
One can see from this representation that most of the field lines in the HFT connect the end and mid parts of the strips or, in other words, the areas of stronger and weaker magnetic field.
The field lines are organized in the HFT in such a way that changing footpoints across the middle part of a given ridge causes swiping of the conjugate footpoint along the other ridge between the ends of the strips.
In the case of the quadrupole configuration in each polarity, the ridges connect two magnetic flux concentrations.
In our present model, they similarly connect the flux concentrations of the AR with the respective polar magnetic ``caps''.}

The grey-scaled $Q$ distribution and the red-blue scaled $K$ distribution can be superimposed on $(\phi,\theta)$-plane by using logical multiplication of pixel colors.
Both distributions can then be combined in one colored graph, allowing one to easily
compare these different characteristics of the field line mapping.
To more clearly illustrate the properties of $Q$ and $K$, we have mapped the resulting image on the spherical surface and superimposed several contours of the corresponding photospheric $B_{r}$ component on it.
The resulting image shown in Figure \ref{f7} summarizes the above discussed properties of the field line mapping in the HFT.

To better understand the HFT structure, it is also very instructive to trace the field lines starting at the ridges of the $Q$ distribution.
In accordance with the above discussion, the conjugate footpoints are located not along the other ridge but rather across it and concentrated on a short arc passing through the point of a local maximum in $Q$.
Each of such two sets of field lines forms a mid-surface of the corresponding QSL, which is further called a quasi-separatrix surface (QSS).
These two surfaces intersect along a field line that we label the quasi-separator (Figure \ref{f7}) for the following two reasons.
First, this field line is characterized by a maximum value of the squashing factor $Q$ and it becomes a genuine separator field line in quadrupole configurations when their flux concentrations shrink to point sources \citep{Titov2002}.
Second, it is characteristic of the separator to pass through two magnetic null points of the configuration.
The quasi-separator partly inherits this property in the sense that its footpoints are located near very localized minima of the magnetic field (Figure \ref{f5}).

In general, the presence of such minima in a given configuration is a good indication that the configuration also contains an HFT.
Determining the minima of the photospheric distribution of $|{\bm B}|$ is much easier than analyzing the magnetic connectivity in the above manner.
Therefore, an initial scanning of the photospheric field minima can provide a quick estimate of the presence of HFTs in more complex configurations.
It is necessary, however, to check additionally that the minima found are not due to the coronal magnetic nulls located right above the minima.
This test can be done, for example, by plotting several iso-surfaces of $|{\bm B}|={\rm const} > B_{\rm min}$ to prove whether they do cover the minima and the gradient of $|{\bm B}|$ has the proper direction.
In the above considered configuration, such a test indeed confirms the absence of the coronal nulls above the photospheric field minima.

The difference between the quadrupole case \citep{Titov2002} and our model is mainly in the type of magnetic flux distribution in one of the two bipoles of the configurations.
In the quadrupole configuration, the flux is concentrated in two spots of the same AR, while in our model, it is distributed over large solar hemispheres.
Such a large length scale leads to a significant increase in our model of maximum values of $Q$, which are four orders of magnitude higher than in the quadrupole field.
In all other respects, the magnetic fluxes in both configurations are partitioned by HFTs in a rather similar way.

\subsection{Sheared/twisted magnetic field}
	\label{s:s/t}

\begin{figure}[htbp]
\epsscale{1.0}
\plotone{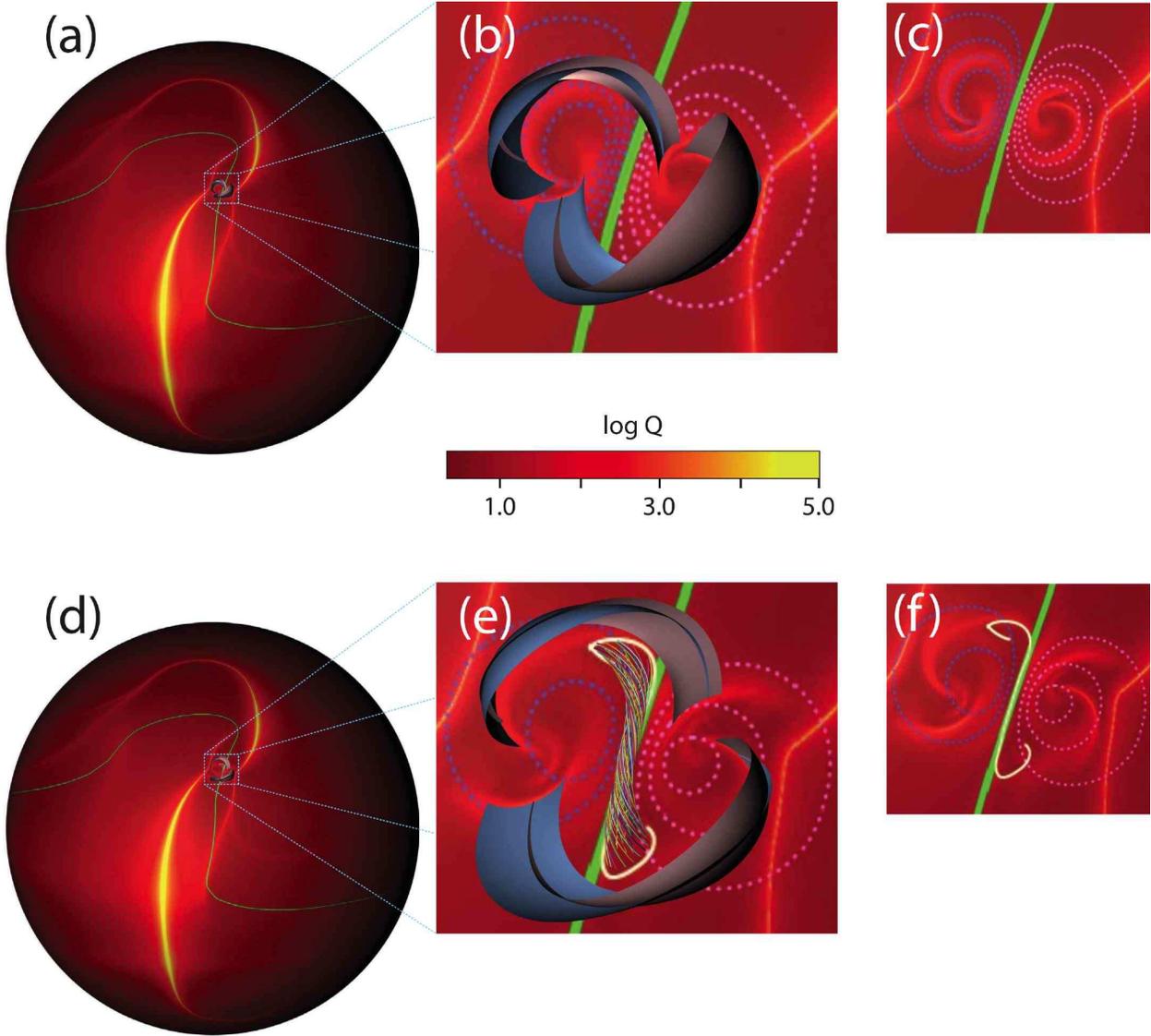}
\caption{Distributions of $Q$ at the end of the shear/twist (a) and flux-cancellation (d) phases; the QSSs in the AR in the first (b) and second (e) cases, respectively, and the corresponding views (c,f) of the $Q$ and $B_{r}$ distributions in the vicinity of the AR; the PIL is a thick green line, the contours of $\left. B_{r} \right|_{r=R_{\sun}} = 44\, i \:{\rm G}$ with $i=-4,\ldots,4$ (a-c) and $i=-2,\ldots,3$ (d-f)  are shown in light blue and pink colors for negative and positive values, respectively; the thick white line in (e,f) is the footprint of the BP SSs wrapping around the flux rope depicted on (e) by a fraction of field lines.
	\label{f8} }
\end{figure}

Figure \ref{f8} (a)--(c) shows the photospheric $Q$ distribution together with some of the new QSSs developed in the configuration after it is subject to the shearing/twisting deformation described in Section \ref{s:sh}.
It can be clearly seen from Figure \ref{f8} that the $Q$ distribution outside the AR changes very little after shearing, which implies a correspondingly small change in the large-scale HFT initially present in the potential field.
However, 
\onecolumn
%
%
\twocolumn
{\parindent=0pt
on a smaller length-scale inside the AR, the magnetic structure changes radically: two new HFTs form at the border of the region that is most affected by the photospheric twisting motions.
These HFTs have modest maxima of $Q$ ($\sim 10^{3}$) compared to the initial large-scale HFT, but they are much shorter and located in the region of largest magnetic field strength.  
Therefore these newly formed small-scale HFTs play a more important role in the flux-cancellation phase than the initial large-scale HFT.
Such an expectation is fully confirmed by the analysis of magnetic structure in the flux-cancellation phase (see sections \ref{s:sfc} and \ref{s:imp}).

The formation of HFTs by a twisting motion is a purely 3D effect that has no analogue in the two-and-half dimensional case of translationally invariant flux tubes. 
%

\begin{figure}[htbp]
\epsscale{1.0}
\plotone{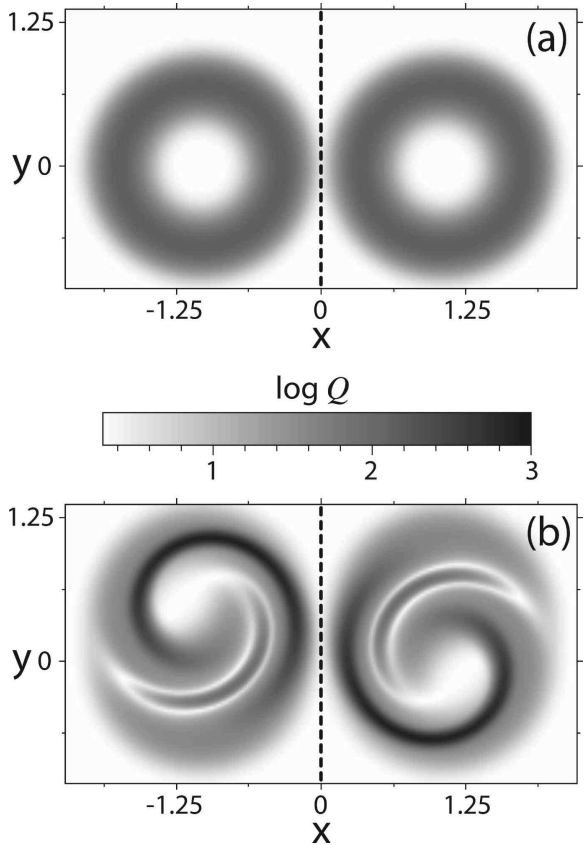}
\caption{Distributions of $Q$ produced by concentric (a) and eccentric (b) vortices in the opposite photospheric polarities at $t=4.0$ with $\Omega=e^{-4\rho^{4}}$, where $\rho$ is the distance from the centers ${\bm r}_{1}$ and ${\bm r}_{2}$ of the vortices.  In both cases ${\bm r}_{1}=(1.05, 0)$, while ${\bm r}_{2}=(-1.05, 0)$ in case (a), and ${\bm r}_{2}=(-1.05, 0.4)$ in case (b).
	\label{f9} }
\end{figure}
%
The vortex motions defined by equation (\ref{v_sh}) and applied to opposite flux spots of the AR have different properties at the conjugate footprints of the twisting flux tube.
In particular, these vortices are not concentric, in the sense that their centers of rotation are not connected by a field line.
In combination with other differences in the vortices, this causes not only twisting but also braiding of the field lines in the coronal volume above the flux spots.
Such a braiding interlocks a part of the field lines and forms the above mentioned small-scale HFTs with less than two windings of the field lines.

In order to describe qualitatively this effect, we construct now a simple analytical example of the field line mapping produced by similar vortices.
The initial field line mapping from positive to negative polarity in the potential configuration can approximately be represented in a Cartesian photospheric plane with the PIL at the $y$-axis by }
\begin{eqnarray}
  {\Pi}_{0} =   \left(\begin{array}{c} x \\
                                          y 
                         \end{array} \right)
   \rightarrow \left(\begin{array}{c} - x \\
                                      \ \ y 
                     \end{array} \right) .
	\label{Fm}
\end{eqnarray}
A circular vortex motion centered at some point ${\bm r}_{\rm c}=(x_{\rm c}, y_{\rm c})$ can be defined as a differential rotation
\begin{eqnarray}
  &\!\!\!\!\!\!\!{\cal R}_{{\bm r}_{\rm c},\, t}&=
    \left(\begin{array}{c} x \\
                           y 
          \end{array} \right)
       \rightarrow
    \left(\begin{array}{c} x_{\rm c} \\
                           y_{\rm c} 
          \end{array} \right) 
           \nonumber \\
           &+&
    \left(\begin{array}{cc}
       \cos\Omega t & -\sin\Omega t \\
       \sin\Omega t & \ \ \cos\Omega t 
          \end{array} \right)
    \left(\begin{array}{c} x-x_{\rm c} \\
                           y-y_{\rm c} 
          \end{array} \right)          
	\label{Rrc}
\end{eqnarray}
acting on a flux spot for time $t$ with angular velocity, say,
\begin{eqnarray}
  \Omega = \exp\left\{-4 \left[ 
      (x-x_{\rm c})^{2} + (y-y_{\rm c})^{2}
   \right]^{2} \right\} .
\end{eqnarray}
The size of the vortex ($\approx 1.2$) corresponds here to the size of the flux spots.
For simplicity, let the vortices be the same in positive and negative polarities and centered at points ${\bm r}_{1}$ and ${\bm r}_{2}$, respectively.
Assuming now that the configuration evolves according to ideal MHD, so that the field line connectivity is preserved, we can compose our field line mapping at time $t$ as 
\begin{eqnarray}
   {\Pi}_{t} = 
   {\cal R}_{{\bm r}_{2},\, t} \circ  {\Pi}_{0} \circ {\cal R}_{{\bm r}_{1},\, -t} .
	\label{Ft}
\end{eqnarray}
This equation means that the conjugate footpoint is obtained in three steps: first, a given footpoint is tracked back in time by ${\cal R}_{{\bm r}_{1},\, -t}$ to get its initial location, from which ${\Pi}_{0}$ provides then the initial location of the conjugate footpoint, whose current location is finally produced by ${\cal R}_{{\bm r}_{2},\, t}$.

Each of the three mappings ${\cal R}_{{\bm r}_{2},\, t}$, ${\Pi}_{0}$, and ${\cal R}_{{\bm r}_{1},\, -t}$ are incompressible, and so is their composition ${\Pi}_{t}$.
This implies that $|\Delta|=1$ in equation (\ref{Q}) and so the squashing factor $Q$  coincides with $N^2$ defined by equation (\ref{N}), where the elements of the Jacobian matrix can easily be obtained from the above definition of ${\Pi}_{t}$.
Figure \ref{f9} represents two calculated examples of $Q$ distributions for symmetric and asymmetric cases, where ${\bm r}_{2}$ and ${\Pi}_{0}({\bm r}_{1})$ are equal and not equal, respectively.
It is natural to call such cases correspondingly concentric and eccentric.
One can see from these examples that even a relatively small ``eccentricity'' in the differential rotations at the opposite polarities causes a substantial effect.
Compared to the concentric case, $Q$ increases here by an order of magnitude when each core of the vortices makes approximately one-half turn.
The larger is the eccentricity in the vortices, the stronger is the squashing of elemental flux tubes at certain layers of the resulting twisted configuration.
This has important implications for the current density distribution in the configuration under study, and also in a more general context, which is further discussed in section \ref{s:imp}.

\subsection{Structure in the flux-cancellation phase}
   \label{s:sfc}

Figure \ref{f8} (d)--(f) shows the $Q$ distribution and QSSs at the final phase of the computation, after flux cancellation has occurred at the PIL.
The small-scale HFTs that are formed after the shearing phase have now increased in size.
This occurs partly because of a noticeable migration of the conjugate footprints outward from the PIL, which is accompanied by reconnection in the HFTs due to a large enhancement of the current density in layer-like structures that will be discussed in section~\ref{s:imp}.

Our simulations confirm also that the photospheric flux-canceling process builds a twisted magnetic flux rope at the PIL \citep{vanBall1989}.
The inverse-S-like footprint of the respective BP SSs appears at this phase in the photospheric $Q$ distribution by outlining the footprints of the flux rope [Figure \ref{f8}(e-f)] in the way predicted earlier for  a similar twisted structure by \citet{Titov1999}.
Here it is evident how universal our technique is---it enables us to identify and locate all the structural elements of the magnetic field under study, including HFTs and BP SSs, as well as flux ropes.
As a matter of fact, the very definition of a flux rope in a given asymmetric 3D configuration can be made precise via this technique.
The flux rope is no more than a fraction of twisted field lines delimited from the surrounding configuration by BP SSs and/or helical QSLs [see another example of the rope in \citep{Titov2007a}].
The presence of such ``delimiters" in the field configuration makes it possible to consider this fraction of field lines as a separate object, namely, a flux rope.

Unfortunately, the boundary conditions chosen for modeling the flux-canceling phase of this run did not produce an eruption, even after reducing the magnetic flux of the AR to half of its initial value.
The fraction of the canceled flux that has been converted into the flux of the rope is not large enough to reach an eruptive unstable state in this particular computational run.
So the evolution of the magnetic structure at the eruptive phase is still to be investigated, which we plan to study after clarifying the eruptive boundary conditions for the flux-canceling process.
We expect that the evolution of the field structure prior to the eruption will remain similar because the HFTs and BP SS we have found are stable structural features that will survive under a wide variety of conditions.

One can anticipate also how the magnetic field structure will change during the initial stage of the eruption process.
Based on the results of \citet{Titov1999}, we expect that detachment of the unstable flux rope will cause first a bifurcation of the BP SS into a pair of BP SSs intersecting along a generalized separator field line.
These bifurcated BP SSs will subsequently transform into an HFT located below the rising flux rope, as demonstrated by \citet{Titov2007a} for a similar equilibrium configuration.
Such a transformation, however, will be a highly dynamical process, pinching this new HFT into a vertical current layer (CL), similar to the one described in previous work \citep{Amari2000, Kliem2004}.
The magnetic reconnection in this CL ought to play a crucial role in making the event eruptive (see sections \ref{s:imp} and \ref{s:s}).

Another important conclusion that follows from our example is that potential and non-potential magnetic fields with the same photospheric flux distributions may have very different structures in both a topological and geometrical sense.
This example undermines therefore a simplifying assumption often used in the analysis of solar magnetic fields, namely that the topologies calculated in potential and non-potential approximations are similar.
\citet{Hudson1999} also argued against this assumption by proposing a counterexample of a constant-$\alpha$ force-free field that has a qualitatively different topology compared to the potential field with the same photospheric normal component of magnetic field.
However, \citet{Brown2000} rightly argued against this counterexample by pointing out that it corresponds to a symmetric configuration, which is topologically unstable to small perturbations of the field.
Their consideration of asymmetric example supported the opposite conclusion that the potential field model generally provides a good approximation for the topology of the observed configurations.
Our configuration is also asymmetric, but it does not support this conclusion, at least for the AR, where it has essentially nonliner force-free field.
This means that the results of topological analysis of realistic magnetic configurations based on the potential approximation can be misleading and should be handled with caution.

\section{FIELD LINE CONNECTIVITY AND CURRENT LAYERS}
	\label{s:imp}

\begin{figure}[htbp]
\epsscale{1.0}
\plotone{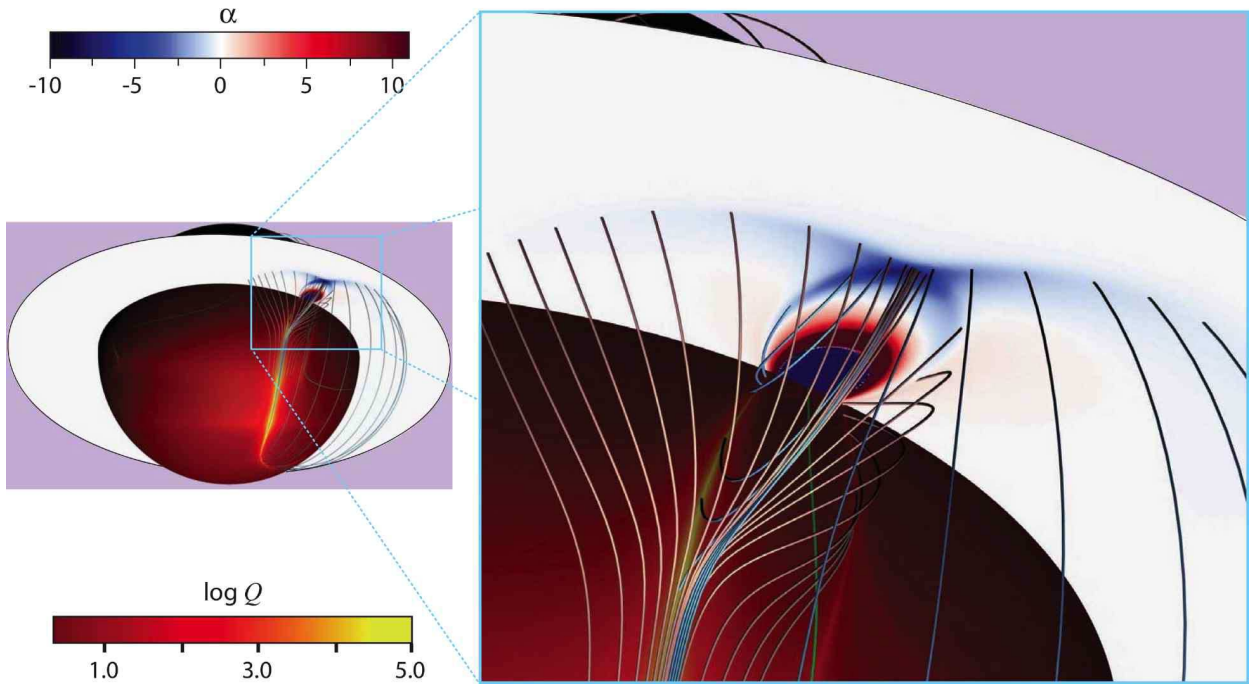}
\caption{The distribution of $\alpha=j_{\|}/B$ (normalized to $1/R_{\sun}$) in the middle cross-sectional plane of the large-scale QSSs (depicted by several field lines) demonstrates an accumulation of the current density perturbations in the large-scale HFT at the end of shear/twist phase.  The corresponding photospheric distribution of $Q$ and the PIL are the same as in Figure \ref{f8}a.
	\label{f10} }
\end{figure}

Photospheric shearing and twisting motions inflate the magnetic field of the AR, which presses up against the overlying large-scale HFT.
The HFT footprints, however, being mostly outside of the applied photospheric vortices, are essentially not affected by these motions, so the line-tied HFT is stretched and pinched to counter the magnetic pressure from the inflating AR field.
This process implies an accumulation of the current density in the HFT such that its X-type mid-cross-section is noticeably pinched in the vertical direction into a \Ht-like shape with a relatively strong CL at the center (Figure \ref{f10}).
The numerical grid used in our MHD code does not enable us to resolve the resulting thin CL and its four adjoining branches in its neighborhood.
Although the cell size of the grid in this region $~5\times10^{-3}\, R_{\sun}$ is several times smaller than the apparent thickness of the CL, it is still much larger than the characteristic thickness of the corresponding QSL (see Figure \ref{f6}), estimated as $R_{\sun}/\sqrt{Q_{\rm max}}  \approx 10^{-5} R_{\sun}$ \citep{Titov2002}.
Nevertheless, the asymptotic tail distribution of such a CL is consistently retrieved by the code with an evident correlation between local maxima of squashing factor $Q$ and parameter $\alpha=j_{\|}/B$ at both shearing/twisting and flux-canceling phases.
A similar correlation has been found by \citet{Aulanier2005} in MHD simulations of the quadrupole magnetic configuration in Cartesian geometry, where the corresponding HFT has been perturbed by distorting its  footprints rather than inflating the underlying structure as in our case.
This is consistent with the structural similarity of our model and the quadrupole configuration as discussed in section \ref{s:sipf}.

One can expect therefore that this CL would lead to reconnection if eruptive conditions were met at the cancellation phase, to produce a rapidly rising flux rope that would push up more strongly against the overlying field structure.
Such a CL is an essential element of eruption models in quadrupole configurations with an emerging inner bipole without shear \citep{Syrovatskii1982} or with strong shear \citep{Antiochos1999}, or with an embedded flux rope \citep{Mackay2006a, Mackay2006b}, as in our case.
We will argue in section
\onecolumn
\twocolumn 
{\parindent=0pt
\ref{s:s} that the reconnection in this CL imposes an important constraint on the eruption process in the event under study.}

\begin{figure}[htbp]
\epsscale{0.8}
\plotone{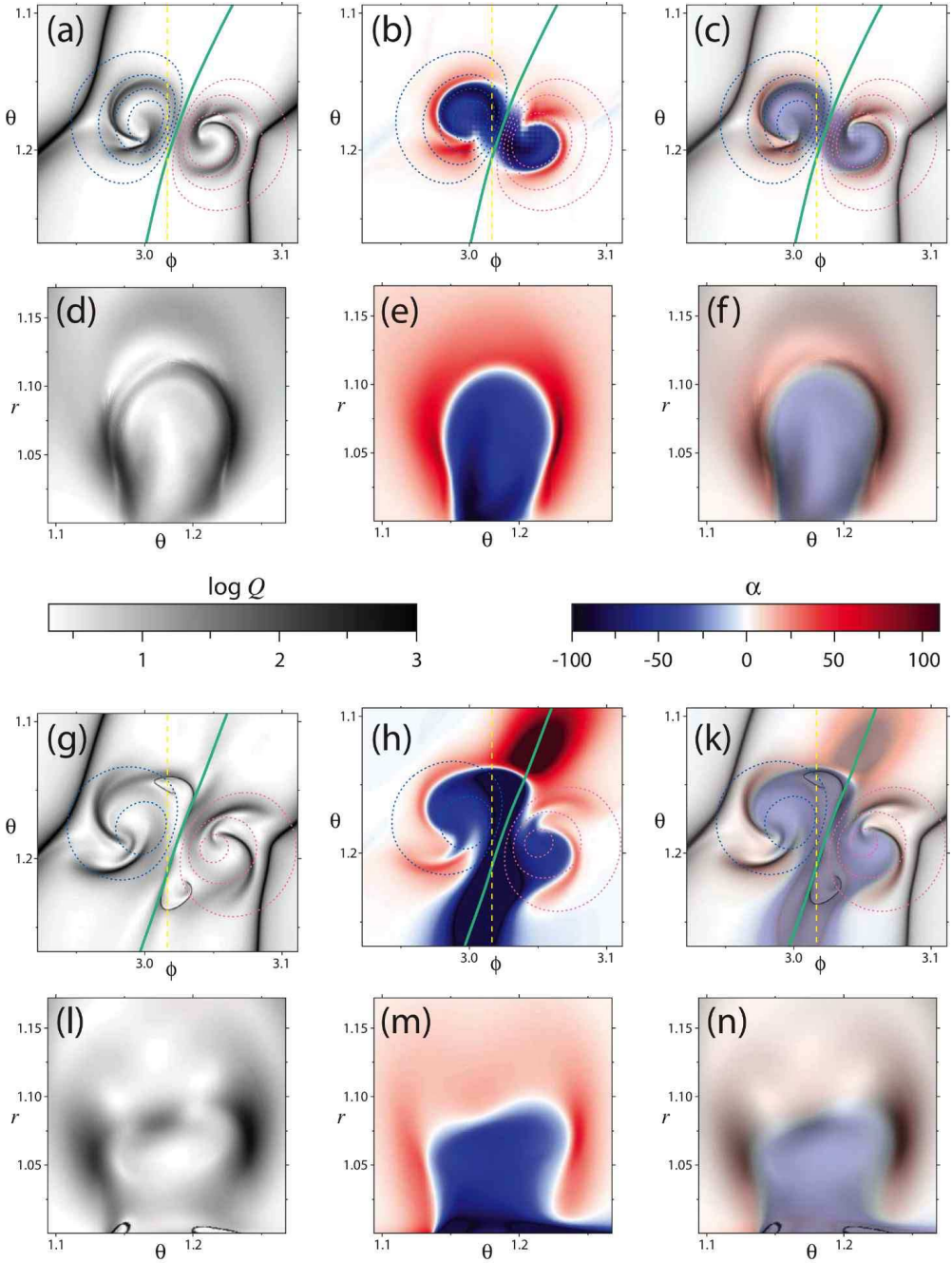}
\caption{Distributions of $\log Q$ [(a), (d), (g), (l)] vs. $\alpha=j_{\|}/B$ [(b), (e), (h), (m)] in the vicinity of flux concentrations at the end of shear/twist [(a)--(f)] and flux-cancellation [(g)--(n)] phases ($\alpha$ is normalized to $1/R_{\sun}$);
panels (a)--(c) and (g)--(k) show these distributions at the photospheric $(\phi,\theta)$-plane, while panels (d)--(f) and (l)--(k) show them at the $(\theta,r)$-plane at $\phi \approx 3.017$ (yellow dashed line in the photospheric distributions).
Superposition of  the corresponding semi-transparent $\log Q$ distributions and $\alpha$ distributions is depicted by panels (c), (f), (k), (n).
The PILs and contours of $\left. B_{r} \right|_{r=R_{\sun}}$ on the photospheric distributions are the same as in Figure \ref{f8}.
	\label{f11} }
\end{figure}

As discussed in section \ref{s:s/t}, two small-scale HFTs are formed in the AR during the shearing/twisting phase and they survive during the flux-cancellation phase of our simulations [panels (a)--(c) and (d)--(f) of Figure \ref{f8} for the former and latter phases, respectively].
In both phases, these HFTs carry strong layer-like currents, as evidenced by the sharp angles at which the corresponding QSSs intersect each other (see Figures \ref{f8}b and \ref{f8}e). In current-free HFTs such angles would be close to right angles.
Figure \ref{f11} shows an evident correlation between local maxima of $Q$ and $\alpha$ and therefore confirms that thin CLs are indeed formed along the HFTs.
One can also see from this figure that $\alpha$ is positive in the CLs and that they are located at the border of the twisted AR field, with negative $\alpha$ in the bulk volume.
So these CLs carry a significant part of the return current that shields the twisted AR field from a nearly potential background field.
Their formation occurs in the coronal volume due to the interlocking of the field lines by a twisting pair of shearing motions eccentrically applied to the conjugate footpoints.
In essence, this is another natural realization of Parker's mechanism of current layer formations discussed by \citet{Titov2003} and \citet{ Galsgaard2003} for quadrupole configurations.
This process has been called magnetic pinching.

Comparing the shearing/twisting and flux-cancellation phases [panels (a)--(f) and (g)--(n), respectively, in Figure \ref{f11}], we see that the CLs increase in size in the latter phase, where a noticeable migration of their footprints outward from the PIL takes place.
Since the magnetic Reynolds number based on the local velocities and length scale is not smaller than $\sim 50$, this migration is likely because of the reconnection that occurs in the CLs rather than due to magnetic diffusion at the photospheric boundary.
Such CLs help redistribute magnetic field and current within the AR at the flux-cancellation phase.
More detailed investigation of this process is required, especially for the case when eruption occurs.

It is clear, however, that its significance extends far beyond this particular simulation.
Indeed, magnetic sunspots with twisting-type motions are quite natural and are often observed on the Sun \citep{Brown2003}.
For this reason, such motions have been considered for a long time as an in situ mechanism to build up magnetic energy in the solar corona.
Our example suggests, however, that the differential rotation of sunspots may also release this energy in the accompanying CLs by magnetic reconnection triggered when the pinching of HFTs reaches a critical level.

It should also be noted that the largest values of $\alpha$ are obtained in our simulations along the flux rope throughout its cross-section (panels (g)--(n) in Figure \ref{f11}) rather than in the CLs. 
There are two reasons for this: first, the perturbation of the initial configuration by shearing/twisting motions may not be sufficiently high to form stronger CLs, and second, these CLs are not well-resolved in our simulations.
The latter may also be the reason why a CL is not formed along the BP SS that wraps around the flux rope, although the formation of such a CL is generally expected in this configuration \citep{Titov1999, Low2003}, even if it evolves quasi-statically.
This is consistent with the numerical simulations of \citet{Fan2004}, where such a CL is formed along the BP SS only in the dynamical kink-unstable phase of the evolution of a similar configuration.  
An additional separate study is required to clarify these issues.


\section{SUMMARY AND DISCUSSION}
	\label{s:s}

We have developed a simplified model of the coronal magnetic field prior to the CME on May 12 1997 in the following three steps.
First, this field is constructed by superimposing a global background field ${\bm B}_{\sun}$ and a localized bipolar field ${\bm B}_{\rm AR}$ to model the active region (AR) in the potential approximation.
The background field ${\bm B}_{\sun}$ is determined from the observed photospheric normal field averaged over the longitude of the Sun.
The AR field is modeled by a subphotospheric dipole whose parameters were optimized to fit the magnetic field obtained from an MDI magnetogram.
To verify the impact of the solar wind on the field structure, ${\bm B}_{\sun}$ is calculated for two different sets of boundary conditions.
In the first case, ${\bm B}_{\sun}$ is modeled as a closed potential field vanishing at infinity.
In the second case, ${\bm B}_{\sun}$ is constrained to be strictly radial and therefore open at the sphere of radius $r=2.5\, R_{\sun}$.
The resulting field structures
\onecolumn
\twocolumn
{\parindent=0pt
in the AR and its neighborhood are quite similar for both choices of boundary conditions, therefore only the first (closed) potential configuration has been used in the next step of MHD modeling.
In this second step, the potential field is quasi-statically sheared by photospheric vortex motions applied to the two flux concentrations of the AR.
Finally, the resulting force-free field is evolved by canceling the photospheric flux with the help of an appropriate tangential electric field applied to the central part of the AR.}

To investigate the structure of the modeled configurations, the field line mapping technique has been generalized to the case of spherical geometry.
In our potential field configuration, this technique reveals a large-scale hyperbolic flux tube (HFT) characterized by very large values of the squashing factor.
As in quadrupole-type configurations, such an HFT consists of two
intersecting quasi-separatrix layers (QSLs) that provide an approximate partition of magnetic fluxes between four photospheric field sources.
In our case, the role of such sources is played by two flux concentrations of the AR and two photospheric polar regions.

It is important that no other structural features, such as magnetic null points or ``bald patches'', are found in the AR and its surrounding area of the initial potential configuration.
In this respect, our initial configuration drastically differs from that used in the breakout model \citep{Antiochos1999}, where the null points and separator are key topological elements of the structure.
On the other hand, our large-scale HFT is a generalized analogue of the separator field line \citep{Titov2002} and is therefore a favorable site for current layer (CL) formation \citep{Titov2003, Galsgaard2003} and magnetic reconnection \citep{Hornig2003a, Aulanier2005, Pontin2005}.
The accumulation of current density observed in the vicinity of this HFT during the shearing/twisting phase of our simulations is in agreement with this point of view.
The mid X-type cross-section of the HFT pinches in the process of this accumulation into a \Ht-like shape with a strong CL at the center.

This structure is preserved during the cancellation phase when the flux rope is formed.
Had the eruption conditions been met in this phase, the flux rope that would form would most likely detach from the dense photosphere and rise by pushing the overlying field upward.
This would cause an additional pinching of the CL to subsequently trigger reconnection within it.
Such a reconnecting CL is a basic element of the breakout model \citep{Antiochos1999} and we have demonstrated that it appears in the flux-cancellation model as well if the interaction of the AR with the global background field is included.
With this generalization, the only difference between the models is in their interpretation of the drivers of the eruption.
The breakout model assumes that the driver is reconnection in the CL, allowing for an explosive expansion of the sheared arcade of the AR. 
The flux-cancellation model assumes instead that the driver is a rising flux rope unleashed from the strapping field of the AR.
While the reconnecting CL is important in our generalized model, the latter mechanism appears to be more likely to drive the eruption. 

%
The following estimate of the magnetic fluxes in the interacting parts of our configuration suggests a picture of how this occurs.
The flux associated with the background field is eight times larger than the initial flux in our AR.
However, only certain fractions of these fluxes interact with each other.
These fractions can be estimated from the geometrical properties of the quasi-separator shown in Figure \ref{f7}.
Since the longitudinal extension of the quasi-separator is about 0.55 radians (Fig. \ref{f5}b), the fraction of the background magnetic flux overlying the quasi-separator is equal to $8\times 0.55/(2\pi)\approx 0.7$. 
We must take into account also that only field components perpendicular to the quasi-separator can participate in reconnection (reducing the fraction of the overlying flux from 0.7 to 0.4), and that the flux of the AR is reduced to half of its initial value at the end of the flux-cancellation phase.
With these corrections, we obtain the interacting fractions of the background and AR fluxes expressed by the ratio $\Phi_{{\rm i}\sun} : \Phi_{{\rm i\, AR}} = 0.4 : 0.5 $.
In other words, the interacting fluxes of the background and AR field are comparable.
It implies that a substantial part of $\Phi_{{\rm i\, AR}}$ must be converted into the magnetic flux associated with the rising unstable flux rope in order to allow it to break through the overlying background field.
Such a conversion is likely provided by magnetic reconnection in the ``vertical'' CL below the rope, as observed in previous simulations by \citet{Amari2000} and \citet{Kliem2004}.
This process ought to replenish the magnetic flux in the rope that is gradually destroyed by reconnection in the ``horizontal" CL above the rising rope.

We have also demonstrated that two small-scale HFTs pinched into thin CLs are formed by nonuniform vortex motions applied to the flux spots of the AR.
The reconnection in these CLs helps redistribute the magnetic flux between the AR and the surrounding field during the flux-cancellation phase.
Other possible implications of their presence should be explored in a more accurate model of the magnetic field evolution in this event.
However, beyond its manifestation in this particular event, the discovered process of current layer formation by rotation of the flux spots has a broader significance.
This process seems to be very important for understanding other eruptive/flaring events driven by rotating sunspots \citep{Brown2003}.
Our example shows that their rotation can serve as a mechanism of both the storage of free magnetic energy and its release via gradual formation of the above CLs in the solar corona.
More detailed studies are certainly required in this direction.

We have also confirmed that the photospheric flux-canceling process builds a magnetic flux rope at the polarity inversion line \citep{vanBall1989}.
This flux rope appears as a well-defined element of the structure owing to the bald patch separatrix surface that wraps around the rope.
The magnetic structure near the rope is topologically equivalent to the twisted configuration described by \citet{Titov1999}.
In particular, the bald patch separatrix surface has a characteristic inverse-S-like shape.
This type of sigmoid corresponds to the negative sign of the current helicity in the flux rope, which is in accordance with observations \citep{Liu2004, Pevtsov1997}.


\acknowledgments

V.S. Titov is thankful to Bernhard Kliem for several useful discussions on magnetic eruptions.
This research was supported by NASA and the Center for Integrated
Space Weather Modeling (an NSF Science and Technology Center).


\appendix

\section{PARAMETERS OF THE FIELD MODEL}
	\label{s:pars}

Table \ref{t1} presents the coefficients of spherical harmonics for
potential background magnetic fields ${\bm B}_{\sun}$ with two
different types of boundary conditions.
In both cases $B_{r\sun}$ is prescribed at $r=R_{\sun}\equiv1$ by
defining the Neumann boundary condition (\ref{BC1}) for the harmonic
potential $F_{\sun}$.
In the first case, the additional asymptotic condition at infinity
$\left. B_{\sun} \right|_{r\rightarrow\infty} \rightarrow 0$ is
required to provide a standard exterior Neumann problem for the
Laplace equation with the spherical boundary $r=R_{\sun}$.
In the second case, the Dirichlet boundary condition
$\left. F_{\sun}\right|_{r=R_{\rm ss}}=0$ at
$r=R_{\rm ss}\equiv2.5$ is imposed instead to define a mixed
boundary value problem for $F_{\sun}$ in the coronal volume $R_{\sun}
\le r \le R_{\rm ss}$, which is a so-called PFSS model.

Table \ref{t2} presents the optimized parameters of the magnetic
dipole in spherical coordinates to model the active-region field ${\bm
B}_{\rm AR}$ through the gradient of the respective dipole potential
(\ref{Far}).


\begin{deluxetable}{rrrrr}
\tablecolumns{5} 
\tablewidth{0pc} 
\tablecaption{Coefficients of spherical harmonics for two types of  $F_{\sun}$ 
	\label{t1}}
\tablehead{ 
\colhead{}    &  \multicolumn{1}{c}{$\left. B_{\sun}
\right|_{r\rightarrow\infty} \rightarrow 0$}   &   \colhead{}
& \multicolumn{2}{c}{$\left. F_{\sun}\right|_{r=R_{\rm ss}}=0$
	\quad $(R_{\rm ss}\equiv2.5 R_{\sun}$)   }   \\ 
\cline{2-2} \cline{4-5} \\ 
\colhead{$n$} & \colhead{$C^{-}_{n}$}  & \colhead{}    
&\colhead{$C^{-}_{n}$}  &\colhead{$C^{+}_{n}$}    }
\startdata 
1	& $1.92227302$	&	&$1.86264984$	&$ -0.119209589$\\ 
2	& $-2.690655399\times10^{-2}$ 	&	&$-2.672416196\times10^{-2}$	&$2.73655421\times10^{-4}$\\ 
3	& $1.33104612$		&&$1.32943307$		&$-2.17814313\times10^{-3}$\\ 
4    	& $-4.470680688\times10^{-2}$	&&$-4.46974778\times10^{-2}$	&$1.171717562\times10^{-5}$\\  
5	& $0.480835203$		&&$0.480804328$  		&$-2.01663951\times10^{-5}$\\ 
6	& $-2.583133904\times10^{-2}$	&&$-2.583117439\times10^{-2}$	&$1.733500769\times10^{-7}$\\ 
7	& $1.1998345\times10^{-2}$	&&$ 1.1982269\times10^{-2}$	&$-1.2865865\times10^{-8}$\\ 
8	& $-5.299226134\times10^{-3}$ 	&&$-5.299194276\times10^{-3}$	&$9.103946449\times10^{-10}$\\ 
9	& $-7.47279826\times10^{-2}$ 	&&$-7.47140767\times10^{-2}$	&$2.0537249\times10^{-9}$\\ 
10	& $-1.633068626\times10^{-7}$  	&&$-1.540580563\times10^{-7}$	&$6.775544972\times10^{-16}$\\  
11	& $-2.40520137\times10^{-2}$  	&&$-2.404263019\times10^{-2}$	&$1.691849693\times10^{-11}$\\
\enddata 
\end{deluxetable} 
\begin{deluxetable}{rcc} 
\tablecolumns{3} 
\tablewidth{0pc} 
\tablecaption{Optimized parameters of  the dipole 
	\label{t2}} 
\tablehead{ 
\colhead{}    &  \multicolumn{1}{c}{$\left. B_{\sun}
\right|_{r\rightarrow\infty} \rightarrow 0$}  
& \multicolumn{1}{c}{$\left. F_{\sun}\right|_{r=R_{\rm ss}}=0$}  
    }
\startdata 
$r_{\rm d}$	& $ 0.9598074306$			&$0.9595343404$\\ 
$\theta_{\rm d}$	& $1.186114672$ 		&$1.185997083$\\ 
$\phi_{\rm d}$	& $3.025771372$		&$3.025680793$\\ 
$m_{r}$   	& $1.740210202 \times 10^{-3}$		&$1.705788497 \times 10^{-3}$\\  
$m_{\theta}$	& $5.831919762 \times 10^{-3}$		&$5.84995686 \times 10^{-3}$\\ 
$m_{\phi}$	& $1.800868961 \times 10^{-2}$		&$1.811699206 \times 10^{-2}$\\ 
\enddata 
\end{deluxetable} 
%




\clearpage



\end{document}